\documentclass[11pt,letterpaper]{article}

\usepackage[utf8]{inputenc}
\usepackage[T1]{fontenc}
\usepackage{lmodern}
\usepackage{microtype}

\usepackage[margin=1in]{geometry}
\usepackage{setspace}
\usepackage{amsmath, amssymb}
\usepackage{graphicx}
\usepackage{booktabs}
\usepackage{longtable}
\usepackage{array}
\usepackage{calc}

\usepackage[htt]{hyphenat}

\usepackage[colorlinks=true, linkcolor=black, citecolor=black, urlcolor=blue]{hyperref}
\usepackage{natbib}
\title{Architecture Dependent Temporal Observability Under Deployment Interference in Edge Inference Systems}

\author{Akul Swami \and Nikhil Chougule}

\date{\today}

\begin{document}

\maketitle

\begin{abstract}

Edge inference systems are typically evaluated with software-reported
latency collected under controlled conditions. We argue, and demonstrate
empirically, that deployment interference can corrupt not only the
inference timing being measured but the timing observability
infrastructure that measures it, and that the two failures can occur
independently.

We pair software-reported timing with externally observable GPIO
intervals captured by a Saleae Logic Pro 8 logic analyzer on an NVIDIA
Jetson Orin Nano, running the same MobileNetV2 model under two inference
architectures (TensorRT FP16 GPU and ONNX Runtime CPU) across baseline,
light memory pressure, and storage writeback stress. Across 35 paired
capture runs (3500 inference samples) plus 3 storage-stress runs where
external pairing failed (300 software-only samples), we observe three
findings the software-only view does not surface. First, the two
architectures differ not only in mean latency but in distributional
structure: TensorRT baseline is tightly clustered near 1.23 ms with
run-mean standard deviation 15 µs across 500 samples, while ORT CPU
baseline is multimodal with a 31.8 ms run-mean standard deviation across
500 samples. Second, light memory pressure inflates TensorRT P99 from
1.28 ms to 1.61 ms across 2000 samples, while one of five ORT
memory-stress runs collapses into a deterministic 198 ms regime (3.5 ms
standard deviation) rather than uniformly inflating variance. Third, and
most consequential, all three TensorRT storage-stress runs produce
complete software timing logs (100/100 iterations) alongside externally
observable timing failures of three different kinds (full post-marker
collapse, \textasciitilde40 \% transition loss, and complete acquisition
failure), while the runtime reports normal completion in every case.

We do not claim TensorRT is unsafe, software timing is universally
invalid, or that we have identified kernel-level root cause. We claim,
narrowly, that timing observability is itself an interference-sensitive
resource, and that summary statistics from a single timing source can
hide failure modes that an independent external observer makes visible.
\end{abstract}

\section{Introduction}

Edge inference benchmarks treat timing instrumentation as a trusted
observer. Latency, throughput, and tail behavior are read off the
runtime's own software-reported timestamps, and conclusions about
deployment suitability are drawn from those numbers. This is reasonable
under controlled conditions. It becomes a measurement validity problem
when the deployment conditions of interest, memory pressure, storage
writeback, scheduling contention, are precisely the conditions under
which the instrumentation itself may degrade
\cite{mytkowicz2009wrongdata, mizrahi2024observer}.

The systems community has long understood that measurement
infrastructure can perturb what it measures
\cite{malony1992intrusion, mytkowicz2009wrongdata, mytkowicz2007perturbation}.
What is less examined in the edge inference benchmarking literature
\cite{ratul2025acceleration, volter2024benchmarking, jeong2022tensorrtframework, jeong2022parallelization}
is a sharper failure mode: the instrumentation does not just perturb the
value, it loses the event. Software timestamps continue to record
completed inferences while an independent external observer of the same
physical line records nothing. The runtime believes everything is fine.
Summary statistics confirm it. Only an out-of-band view exposes the gap.

This paper reports on a small empirical study designed to make that gap
measurable. We wrap each inference call with a GPIO transition asserted
before execution and de-asserted after completion, capture the resulting
pulse train on a Saleae Logic Pro 8 logic analyzer, and pair each
external pulse with the corresponding software-reported interval using a
long synchronization marker. The setup gives us two parallel timing
streams that should agree under normal conditions and diverge in
informative ways when they do not.

\begin{figure}
\centering
\includegraphics[width=\linewidth,keepaspectratio]{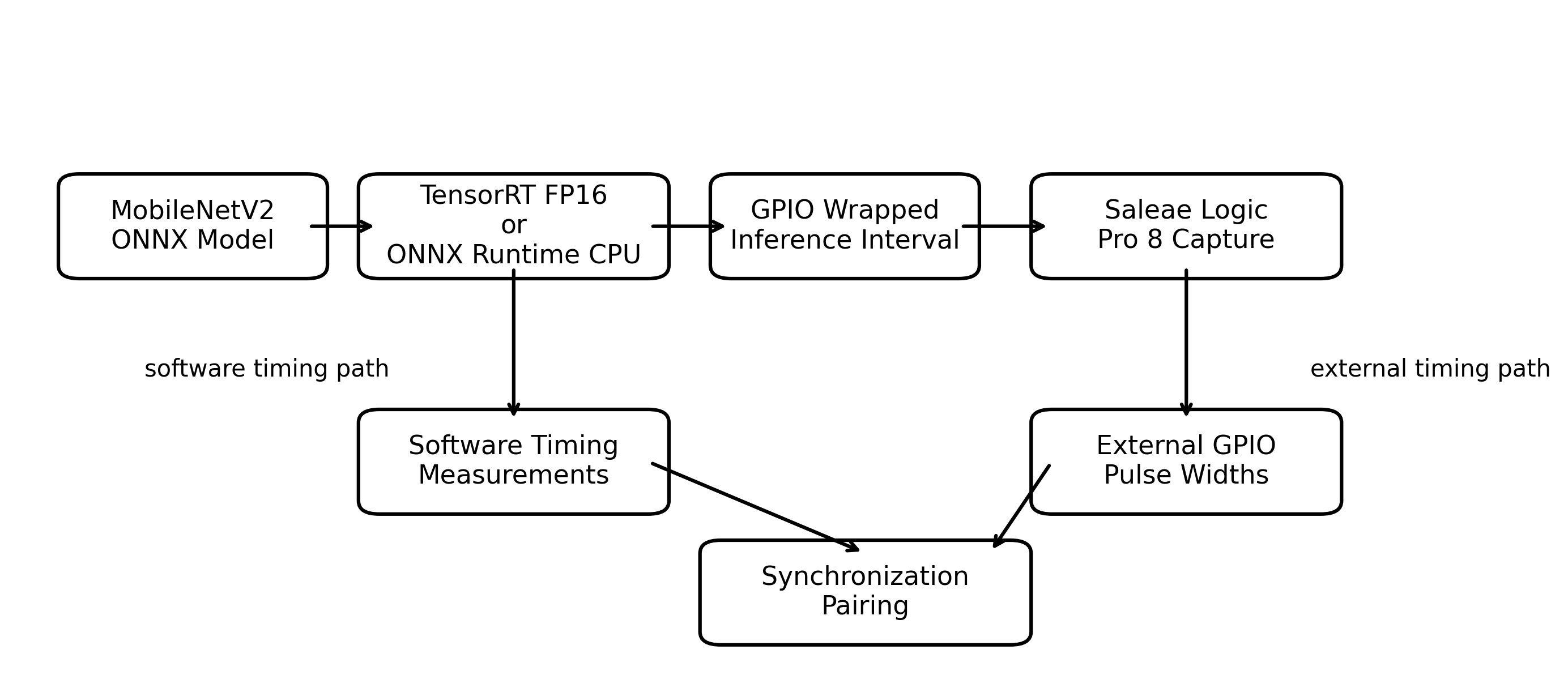}
\caption{Experimental timing validation architecture.}
\label{fig:f1}
\end{figure}

\begin{figure}
\centering
\includegraphics[width=\linewidth,keepaspectratio]{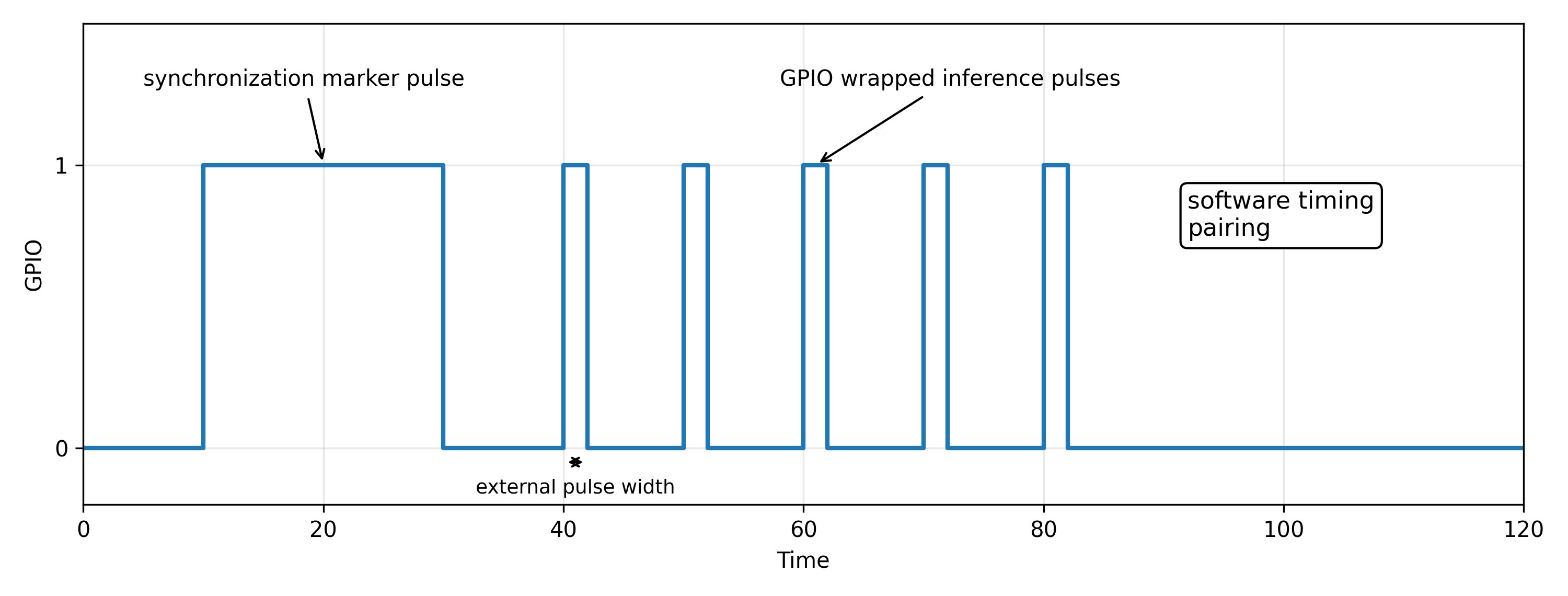}
\caption{GPIO wrapped synchronization methodology.}
\label{fig:f2}
\end{figure}

Our \textbf{contributions} are:

\begin{enumerate}
\def\labelenumi{\arabic{enumi}.}
\item
  A constrained empirical comparison of TensorRT FP16 GPU and ONNX
  Runtime CPU inference on the same Jetson Orin Nano platform under
  baseline, memory pressure, and storage stress, with externally
  synchronized hardware ground truth on every reported timing value
  (Section 4).
\item
  Evidence that deployment interference affects inference architectures
  qualitatively differently. TensorRT tail latency inflates while
  preserving distribution shape; ORT exhibits a regime shift in which
  one of five memory-stress runs collapses into a deterministic slow
  mode rather than broadening variance (Section 4.2--4.3).
\item
  Direct evidence that timing observability can fail independently of
  inference execution. In all three TensorRT storage-stress runs,
  software timing CSVs record 100 of 100 inferences while external
  transition acquisition is degraded in three different ways: full
  post-marker collapse, partial transition loss (\textasciitilde40 \%),
  and complete acquisition failure. The runtime cannot detect any of
  these from its own measurements (Section 4.4).
\item
  Two methodological lessons that emerged from the work and that we
  believe generalize: synchronization markers must be width-separated
  from the inference distribution to avoid silent pairing corruption
  (Section 3.2), and the first measured inference is not the same
  population as steady-state inferences unless warmup is structurally
  enforced (Section 3.3).
\end{enumerate}

We \textbf{do not} claim that TensorRT is unsafe, that GPU inference is
broken, that software timing is universally unreliable, or that we have
identified the kernel-level cause of any observed phenomenon. We claim
only that observability is itself stressed by the deployment conditions
under which inference is most worth evaluating, and that external
corroboration changes which conclusions are defensible.

\section{Related Work}

The phenomenon that measurement infrastructure can perturb what it
measures is foundational in systems performance analysis. Malony, Reed,
and Wijshoff \cite{malony1992intrusion} formalized intrusion and
perturbation in the early 1990s, showing that tracing overhead can
exceed three orders of magnitude in execution time and that the
resulting timing distributions cannot be trusted as ground truth.
Mytkowicz and colleagues
\cite{mytkowicz2009wrongdata, mytkowicz2007perturbation} later sharpened
the result for modern systems, demonstrating that even sub-3\%
instrumentation overhead suffices to alter measured behavior in ways
that propagate through entire performance studies, and that a survey of
systems papers in this era largely failed to account for this.
Tracing-overhead studies on HPC clusters \cite{mohror2006tracing} and
real-time testing analyses \cite{schutz1994realtime} reach compatible
conclusions in their respective domains: the observer is part of the
system. Mizrahi, Schapira, and Moses \cite{mizrahi2024observer} recently
revived the framing for computer networks, where measurement accuracy
and overhead are explicitly traded against each other.

The edge inference benchmarking literature comes at deployment
characterization from a different direction. Ratul, Zhou, and Yang
\cite{ratul2025acceleration} compare five frameworks (PyTorch, ONNX
Runtime, TensorRT, TVM, JAX) on Jetson AGX Orin, treating
software-reported latency and throughput as the primary measurement.
Völter, Koppe, and Rieger \cite{volter2024benchmarking} propose
tritonPerf as a pre-deployment benchmarking framework, again building on
software-side timing. Jeong and colleagues
\cite{jeong2022tensorrtframework, jeong2022parallelization} focus on
optimization frameworks and heterogeneous-processor parallelization for
TensorRT on Jetson, where benchmark numbers drive design decisions.
These works are valuable and we rely on the framework characterizations
they provide. They are not, by design, asking whether the
instrumentation generating those numbers remains reliable under
interference.

Our work sits at the intersection. We are not proposing a new inference
framework or scheduler. We are not formalizing perturbation theory. We
are pointing out that the systems-observability concerns from
\cite{malony1992intrusion, mytkowicz2009wrongdata, mytkowicz2007perturbation, mohror2006tracing, schutz1994realtime, mizrahi2024observer}
apply directly to the edge-inference benchmarking practice exemplified
by
\cite{ratul2025acceleration, volter2024benchmarking, jeong2022tensorrtframework, jeong2022parallelization},
and that the combination produces empirically observable failure modes
that an external timing channel can detect and an internal one cannot.
The contribution is empirical: we show what those failure modes look
like on real hardware with real workloads, and what conclusions a
single-timing-source study would have drawn instead.

\section{Experimental
Methodology}

\subsection{Platform and
workloads}

All experiments ran on an NVIDIA Jetson Orin Nano (L4T R36.5.0, Linux
5.15.185-tegra, CUDA 12.6, TensorRT 10.3.0), with a Saleae Logic Pro 8
logic analyzer on Jetson BOARD pin 29, sampled at 10 MS/s with a 1.8 V
logic threshold. We used a single MobileNetV2 ONNX model (input {[}1, 3,
224, 224{]}, output {[}1, 1000{]}) compiled to an FP16 TensorRT engine
(7,538,604 bytes, batch size 1) for the GPU path and the same ONNX file
under onnxruntime CPU execution provider for the CPU path. Each capture
run executed 100 inference iterations.

\subsection{External
synchronization}

Each inference iteration asserted GPIO high immediately before the
inference call and de-asserted GPIO low immediately after completion
(after \texttt{cudaStreamSynchronize} for TensorRT, after
\texttt{session.run} return for ORT), following the per-platform
GPIO-wrapping protocol we previously characterized for cross-platform
overhead \citep{swami2026gpio}. Each capture session began with a long
synchronization marker pulse, intentionally configured to exceed the
inference pulse distribution by an order of magnitude. The offline
analysis pipeline classified pulses above an architecture-specific
threshold as markers and pulses below as inference intervals, then
paired software-reported intervals with external pulses by index from
the marker forward. No synchronized clock between the Jetson and the
logic analyzer host was required; correspondence relied on the marker as
a one-time anchor.

\textbf{Marker design lesson.} Our initial ORT configuration used a 200
ms marker, chosen for symmetry with the TensorRT path. The ORT CPU
inference distribution spans approximately 80--250 ms, which overlaps
the marker width. The analyzer misclassified slow inference pulses as
markers and produced silently corrupted pairings. The fix was a 1000 ms
ORT marker with an 800 ms classifier threshold, which is well above the
ORT inference distribution. We report this because the failure was not
visible in any single capture's summary statistics, the pairing looked
superficially valid, and we believe the general lesson (marker width
must be separated from the inference distribution by a margin that
exceeds the distribution's expected stretch under interference) applies
to any externally synchronized capture methodology.

\subsection{Runtime structure and
warmup}

Initial TensorRT runs placed the first measured inference immediately
after the marker pulse. The first sample showed approximately 3.21 ms
while subsequent samples clustered near 1.22 ms. A startup transient
(engine initialization, CUDA context, first-call allocator behavior) was
contaminating the distribution. The corrected runner performs ten
unmeasured warmup inferences before the marker. We report this not as a
fix to a bug but as a structural lesson: the first measured sample is
rarely from the same population as steady-state samples, and ``skip the
first N samples'' applied at analysis time is weaker than enforcing
warmup structurally in the runner. The corrected baseline is what we
report throughout.

\subsection{Stressor methodology}

Stressors were generated with \texttt{stress-ng}. Light memory pressure
used \texttt{-\/-vm\ 1\ -\/-vm-bytes\ 25\%\ -\/-vm-keep}, producing
sustained allocation and paging activity at modest intensity. Storage
writeback pressure used
\texttt{-\/-hdd\ 1\ -\/-hdd-bytes\ 2G\ -\/-hdd-write-size\ 4M},
producing sustained block-layer writeback against the eMMC. Stressors
ran for the full duration of each capture session.

\subsection{Validity
classification}

Following our internal measurement validity protocol, each run was
classified into one of four states: (a) valid runtime \emph{and} valid
synchronization, (b) valid runtime, incomplete synchronization, (c)
invalid runtime, or (d) methodology failure example. Aggregate external
timing claims include only class (a). Runtime-only software timing
claims may include class (b) when the software CSV is complete and the
failure is in the external channel. We do not silently delete class (b)
runs; they are the central evidence for Section 4.4.

\subsection{Statistical
characterization}

For each (architecture × condition) cell we report number of runs, total
samples, per-run mean, run-mean standard deviation, run standard
deviation, per-run P95 and P99, and the maximum observed across runs. We
visualize with ECDFs and run-level boxplots because several conditions
exhibit broad or multimodal distributions where mean and standard
deviation alone are misleading.

\section{Results}

\subsection{Cross-architecture baseline
behavior}

Table 1 summarizes per-condition aggregates across all reported runs.
Figure 3 shows cross-architecture run-level latency profiles at P50,
P95, P99, and Max.

\textbf{Table 1. Per-condition timing summary (run-level aggregates).}

\begin{center}
\small
\begin{tabular}{@{}lrrrrrr@{}}
\toprule
Condition & Runs & Samples & \begin{tabular}{@{}r@{}}Mean of run\\means (ms)\end{tabular} & \begin{tabular}{@{}r@{}}Run-mean\\SD (ms)\end{tabular} & \begin{tabular}{@{}r@{}}Mean P99\\(ms)\end{tabular} & \begin{tabular}{@{}r@{}}Max\\observed (ms)\end{tabular} \\
\midrule
TensorRT baseline                & 5  & 500  & 1.228   & 0.015  & 1.276   & 3.924   \\
TensorRT memory stress (light)   & 20 & 2000 & 1.469   & 0.026  & 1.613   & 6.790   \\
TensorRT storage stress          & 3  & 300  & 1.485   & 0.009  & 2.316   & 2.876   \\
ORT CPU baseline                 & 5  & 500  & 171.043 & 6.644  & 219.504 & 249.560 \\
ORT CPU memory stress (light)    & 5  & 500  & 177.544 & 12.852 & 209.251 & 265.392 \\
\bottomrule
\end{tabular}
\end{center}

\begin{figure}
\centering
\includegraphics[width=\linewidth,keepaspectratio]{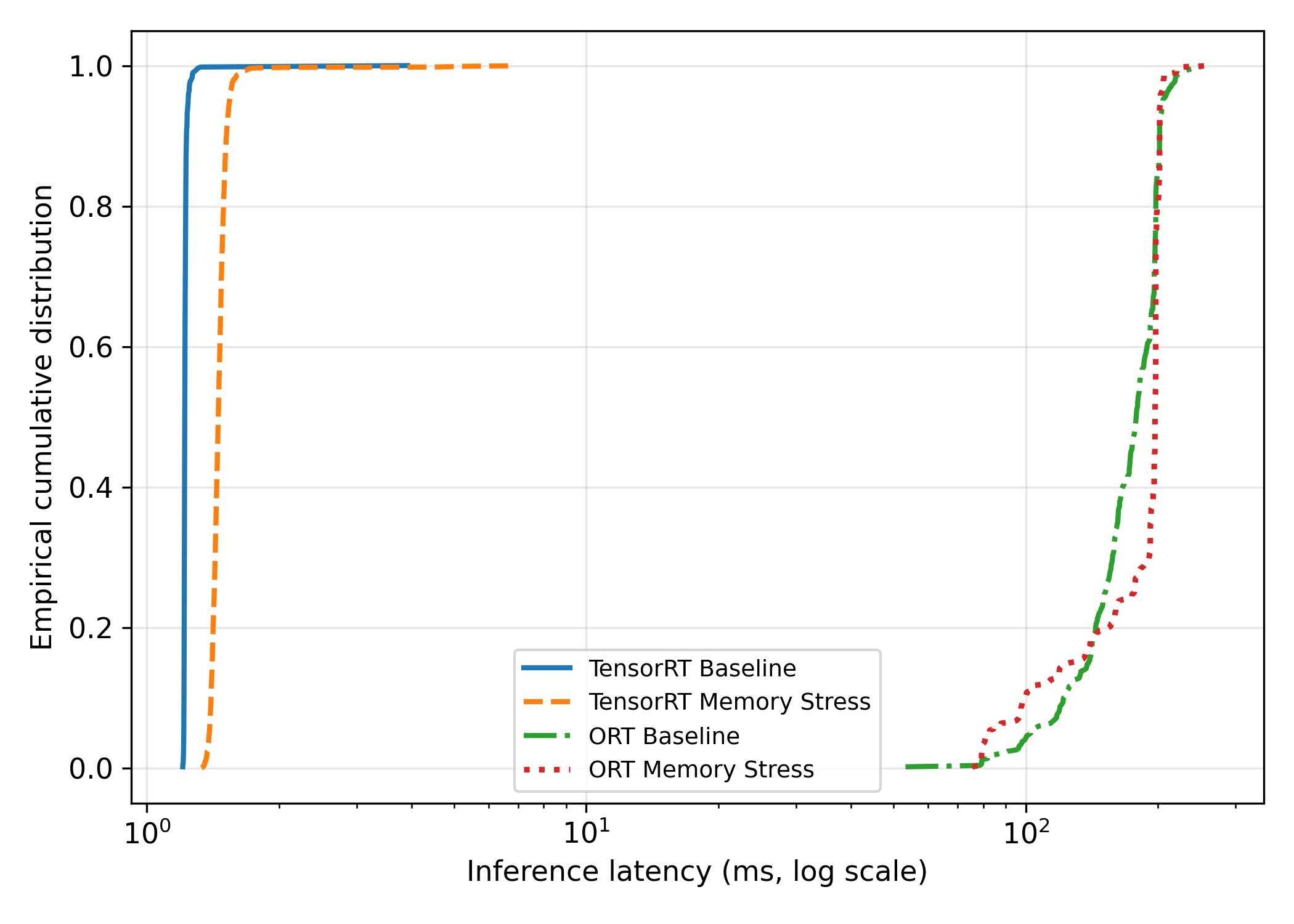}
\caption{Cross-architecture run-level latency profile.}
\label{fig:f3}
\end{figure}

The two architectures differ by roughly two orders of magnitude in
absolute latency on the same hardware, same model, same input. More
relevant to the observability argument: their \emph{distributional
structure} differs as well. TensorRT baseline has a run-mean standard
deviation of 15 µs across 500 samples; the runtime is internally
consistent across capture sessions. ORT CPU baseline has a run-mean
standard deviation of 6.6 ms and includes individual run standard
deviations between 26 and 38 ms, reflecting intrinsic multimodal
behavior visible in per-run histograms (clusters near 80 ms, 120--160
ms, and 190--220 ms across captures).

A study reporting only mean latency would conclude that TensorRT is
\textasciitilde140× faster than ORT, which is true but uninformative for
deployment characterization. The shape of the ORT distribution is the
story.

\subsection{TensorRT under memory
pressure}

\begin{figure}
\centering
\includegraphics[width=\linewidth,keepaspectratio]{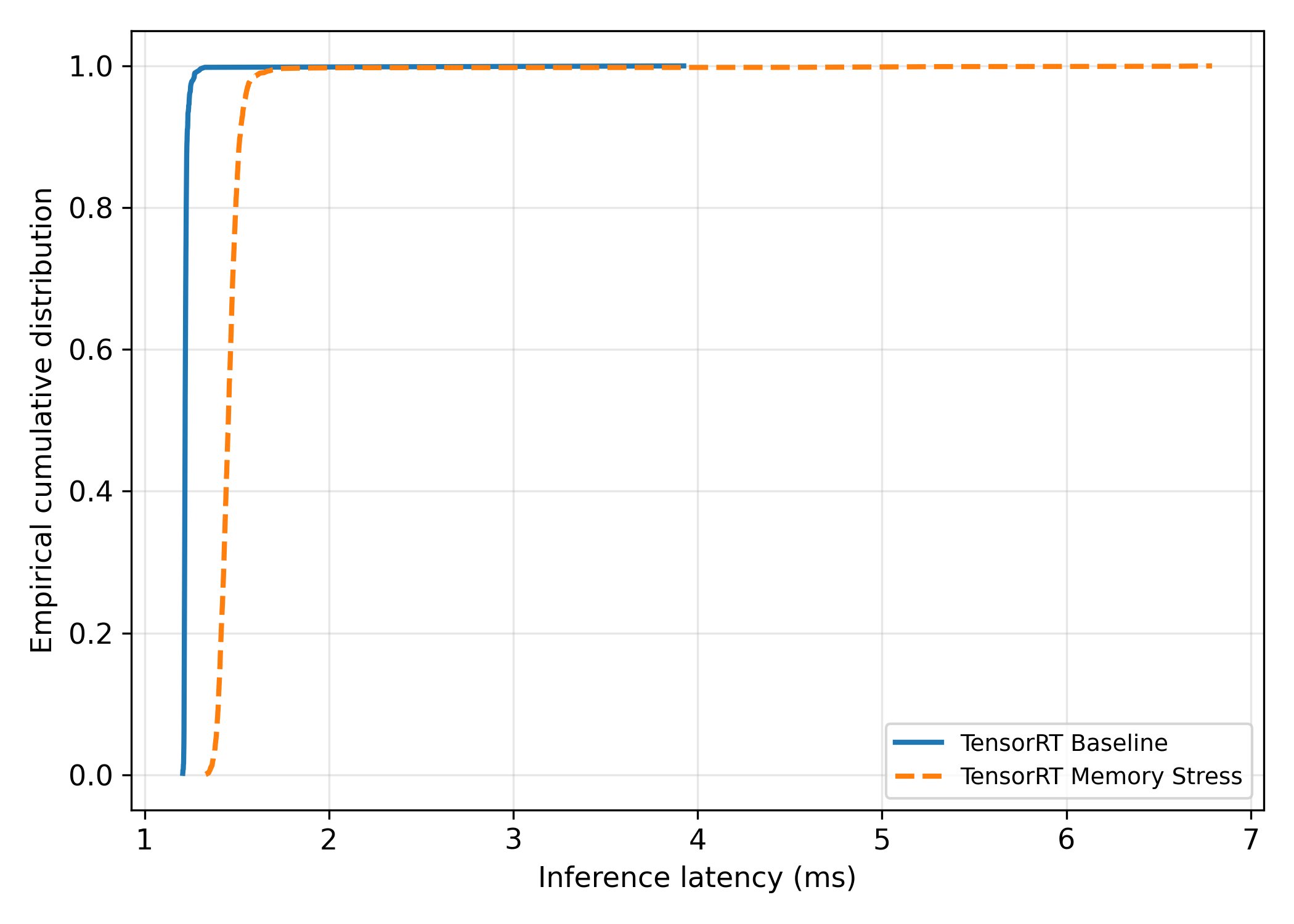}
\caption{TensorRT run-level tail latency under baseline and
memory pressure.}
\label{fig:f4}
\end{figure}

Under \texttt{stress-ng\ -\/-vm} memory pressure, the TensorRT
distribution shifts measurably without changing shape. The mean moves
from 1.228 ms to 1.469 ms (+19.6\%), but the more pronounced change is
in the tail: mean P99 moves from 1.276 ms to 1.613 ms (+26.4\%), and the
worst observed sample across 2000 samples reaches 6.79 ms, more than 5×
the baseline mean. The 20-run memory-stress sample includes four runs
(10, 13, 16, 20) with run standard deviations in the 0.35--0.53 ms range
driven by isolated spikes; the remaining 16 runs sit between 0.035 and
0.078 ms standard deviation. The interference is not uniformly
distributed: most inferences are unaffected, but a small fraction take
dramatically longer.

A summary-statistics view (mean +19.6\%, headline result) would
understate this; a P99 view (mean P99 +26.4\%, max +73\%) is closer to
the deployment-relevant signal. This is consistent with the broader
observation from systems performance work
\cite{malony1992intrusion, mytkowicz2009wrongdata, mytkowicz2007perturbation}
that interference effects concentrate in the tail rather than spreading
uniformly across the distribution.

\subsection{ORT CPU under memory pressure: regime
shift}

\begin{figure}
\centering
\includegraphics[width=\linewidth,keepaspectratio]{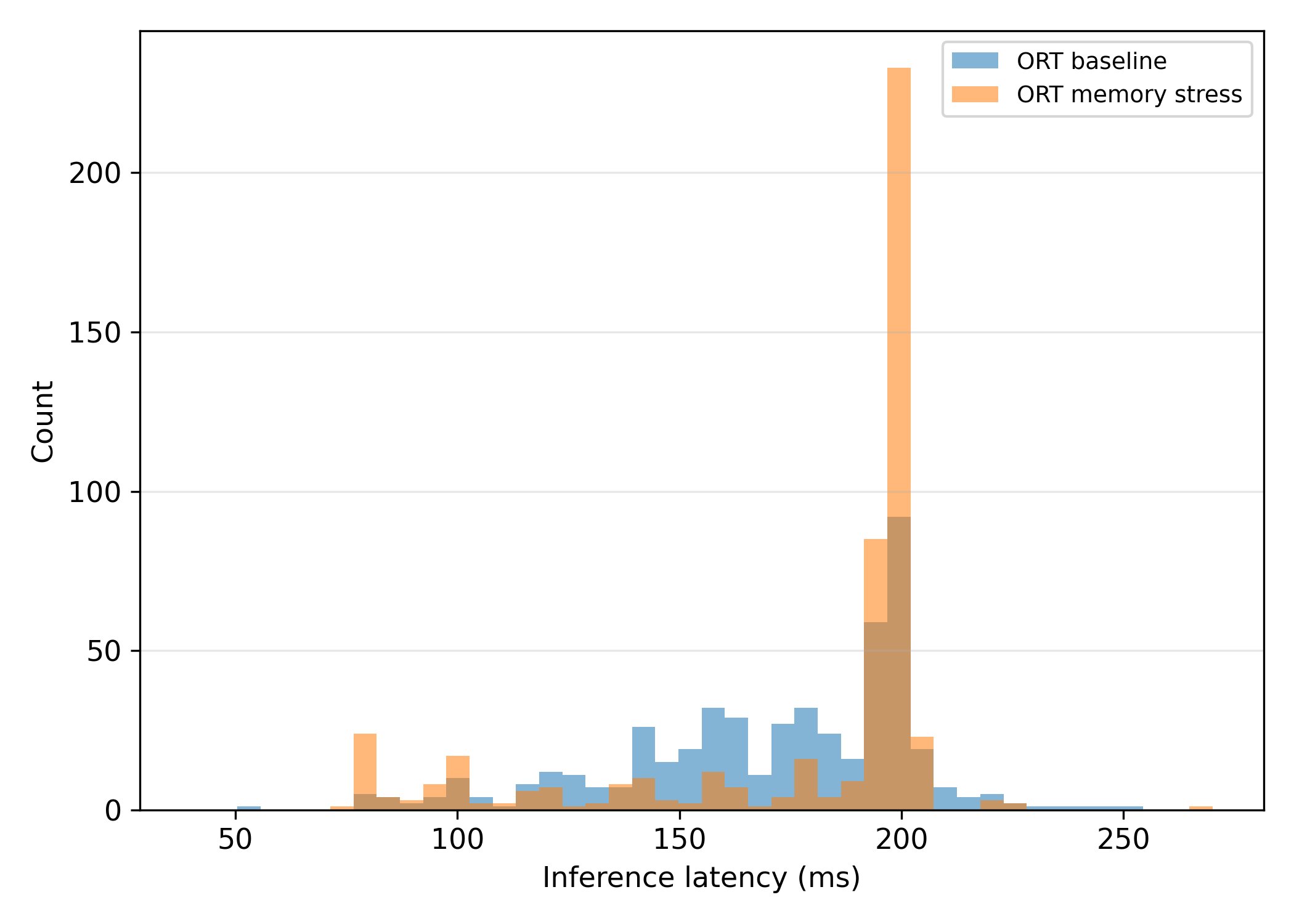}
\caption{ONNX Runtime CPU run-level latency profile under
baseline and memory pressure.}
\label{fig:f5}
\end{figure}

The ORT CPU memory-stress result is qualitatively different. Aggregate
statistics suggest a modest shift: mean from 171.04 ms to 177.54 ms
(+3.8\%), mean P99 actually \emph{decreases} slightly from 219.50 ms to
209.25 ms. This appears reassuring. It is not.

Looking at individual runs, four of the five ORT memory-stress runs
preserve the multimodal baseline structure with run standard deviations
of 34--46 ms. The fifth run (\texttt{ort\_memstress\_light/005})
collapses into a deterministic regime: mean 198.32 ms, standard
deviation 3.50 ms, P95 202.36 ms, P99 205.67 ms, max 205.73 ms. The
distribution is tighter than baseline by an order of magnitude in
spread, but anchored at the slow mode of the multimodal baseline
distribution.

We do not claim to know why this happens; we have not done kernel-level
attribution. The empirical observation is that deployment interference
need not \emph{broaden} a runtime's temporal behavior. It can
\emph{narrow} it onto a slower deterministic regime, in a way that
aggregate statistics across runs may obscure. This is the kind of result
that motivates run-level rather than condition-level reporting.

\subsection{Storage stress: observability decoupling from
execution}

This is the central methodological result of the paper.

\begin{figure}
\centering
\includegraphics[width=\linewidth,keepaspectratio]{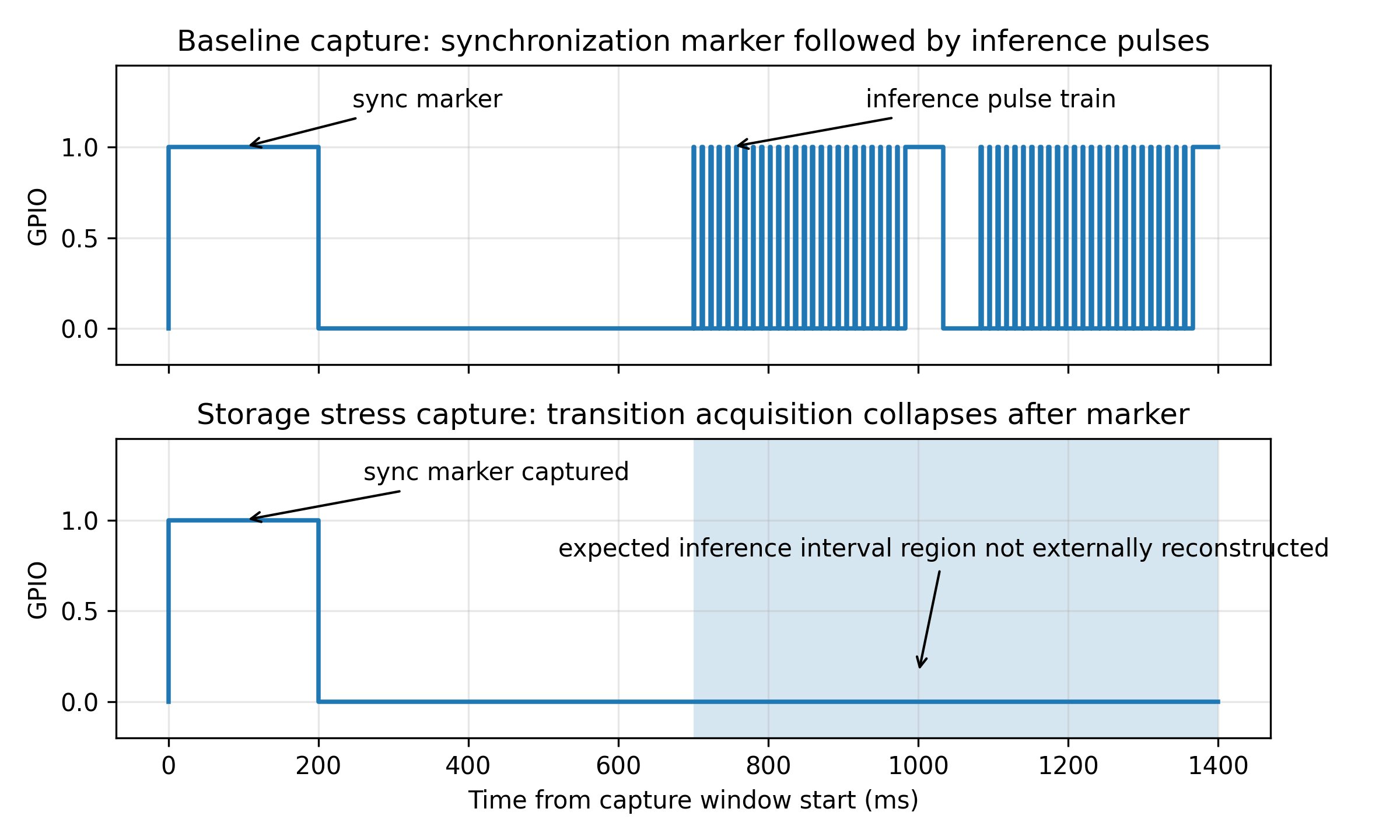}
\caption{Synchronization degradation under storage writeback
stress (Run 001 shown).}
\label{fig:f6}
\end{figure}

Across three TensorRT storage-stress runs, every software timing CSV
recorded 100 of 100 inference iterations with no missing rows, finite
latencies, and a P99 of 2.32 ms (consistent with mild tail inflation).
From the software's perspective, inference was healthy in every run. The
same three runs produced external acquisition outputs that were degraded
in three qualitatively different ways. \textbf{Run 001} (Figure 6)
captured the synchronization marker cleanly and then lost the entire
post-marker inference pulse train, 2 transitions recovered against 200
expected (100 rising + 100 falling). \textbf{Run 002} captured the
marker and approximately 60 of 100 expected inference intervals (122
transitions in total), enough to confirm inference activity but with
substantial pairing loss. \textbf{Run 003} produced an empty acquisition
trace, no marker, no inference pulses, while the software CSV again
reported 100 completed iterations. We verified all three captures at the
raw \texttt{.sal} level to rule out a CSV export pipeline artifact.

We can localize the failure point in runs 001 and 002 because at least
the marker was recovered in both cases; the analyzer was acquiring and
the GPIO line did toggle. Run 003 is more ambiguous: it is consistent
with either Jetson-side GPIO toggle failure (the inference loop
completed but GPIO assertions silently failed) or host-side capture
pipeline failure under storage pressure on the analyzer host. The
software CSV alone cannot distinguish these. We report the ambiguity
rather than choose between explanations.

The three runs do not exhibit a single failure mode. They exhibit a
spectrum: total post-marker collapse, partial collapse, and complete
acquisition failure. The common thread is that in every run, software
timing remained complete while external acquisition was degraded to some
degree. We classify this as observability decoupling: inference
execution continuity (software stream complete) and timing observability
continuity (external stream impaired) are not the same property and do
not fail together. Aggregate external timing claims for storage stress
are not defensible from this dataset, there are not enough paired
samples across the three runs to characterize the latency distribution,
but the runtime would never know. A study reporting only software timing
under storage stress would publish a complete latency distribution and
have no signal that the underlying timing infrastructure had degraded
across all three attempted captures.

We do not know the mechanism, and the three different failure modes
suggest it may not be a single mechanism. Candidate explanations include
block-layer writeback interfering with the GPIO driver's I/O path, eMMC
contention causing transition events to be enqueued past the analyzer's
buffer window, the GPIO toggling itself becoming high-latency in a way
that compresses adjacent transitions below the 10 MS/s sample
resolution, and capture-host storage pressure during the analyzer write
path. Distinguishing these would require kernel-level tracing on the
Jetson and instrumentation of the capture host, neither of which we
undertook. The point of the finding does not depend on the mechanism.
The point is that under a single common deployment stressor (one
\texttt{stress-ng\ -\/-hdd} worker), the hardware-observable timing
channel diverged from the software-observable one in three different
ways across three consecutive runs, and the divergence was invisible
from the software side every time. With n=3, we cannot distinguish a
deterministic failure regime from a rare failure that recurred; we
report the limitation in Section 6.

\subsection{Failure taxonomy}

\begin{figure}
\centering
\includegraphics[width=\linewidth,keepaspectratio]{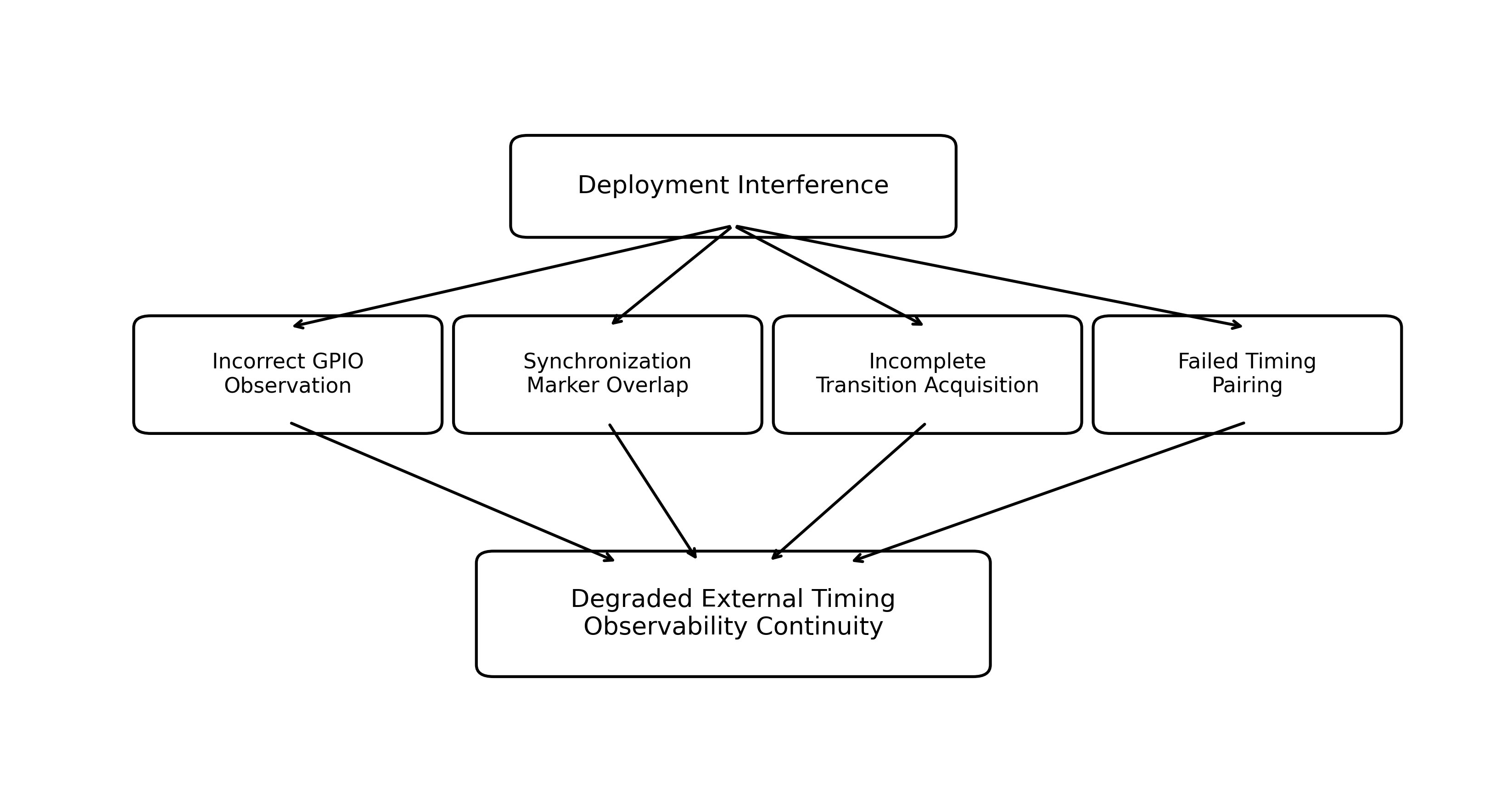}
\caption{Timing observability failure taxonomy.}
\label{fig:f7}
\end{figure}

Across the full set of experiments we encountered four distinct failure
modes that the validation methodology had to handle explicitly: (a)
incorrect GPIO line observation (resolved by physical verification of
pin mapping at setup), (b) synchronization marker overlap with inference
pulses (Section 3.2), (c) incomplete transition acquisition under stress
(Section 4.4), and (d) failed software-to-external pairing for runs with
partial external data. Figure 7 summarizes these as branches of
``deployment interference → degraded timing observability.'' We report
the taxonomy not as a contribution of comprehensive failure-mode
coverage, it is not, but as an honest accounting of what we encountered,
and as a checklist for anyone building externally synchronized timing
methodology on similar platforms.

\section{Discussion}

The experiments support a narrow but specific argument. Different
inference architectures on the same hardware exhibit qualitatively
different temporal behavior, and that difference is not captured by mean
latency. TensorRT and ORT differ in baseline distribution shape; they
differ further in how their distributions respond to memory pressure
(tail amplification for TensorRT, regime shift for ORT); and at least
one common deployment stressor (storage writeback) produces
observability failure that the runtime cannot detect through its own
timing stream.

The methodological implication is that single-source timing measurement
is sufficient for benchmarking under controlled conditions but not for
characterizing behavior under deployment interference. External
corroboration is one way to detect the failure modes we encountered; it
is probably not the only way. Differential consistency checks (do two
independent software timing sources agree?), invariant-based validation
(do reported intervals sum to wall-clock time?), and statistical anomaly
detection on the timing stream are alternatives. Our preference for
external GPIO capture is partly methodological and partly practical: the
channel is fully independent of the runtime, the hardware is
inexpensive, and the failure modes (marker overlap, transition loss) are
themselves informative.

We are explicit about what the experiments do not show. They do not show
that TensorRT is unsafe; the TensorRT path is the better-behaved of the
two by every metric we collected. They do not show that ORT is
unsuitable; ORT's broader distribution is a property of the CPU
execution path and is informative, not disqualifying. They do not show
that software timing is unreliable in general; across the 35 paired
capture runs, the software and external streams agreed to within the
resolution of the analyzer. The divergence we report is concentrated in
the three storage-stress runs, all of which exhibited some form of
external acquisition degradation. The experiments do not establish
kernel-level mechanism for any of the observed phenomena. The
contribution is empirical: under specific conditions, on specific
hardware, the two timing channels diverge, and the divergence carries
deployment-relevant information.

\section{Limitations}

This is a small study on a single hardware platform with one model, two
runtimes, and two stressors. The generalization claims are
correspondingly narrow. We have not characterized network interference,
thermal stress, multi-tenant accelerator scheduling, or sensor-driven
execution; the failure modes under those conditions may look different.

The storage-stress finding rests on n=3 runs. Three observed failure
modes across three consecutive runs is suggestive but cannot distinguish
a deterministic failure regime under storage pressure from a rare
failure that happened to recur three times. A larger storage-stress
sample, and ideally a controlled comparison against a non-stressor
baseline of the same length, would be needed to claim a failure rate. We
report what we observed; we do not claim a rate.

The external GPIO methodology has its own limits. Pulse width measured
at the analyzer includes GPIO toggle overhead, CUDA stream
synchronization, Linux userspace scheduling, and C++ runtime behavior in
addition to the inference kernel itself. We treat the externally
observable interval as ground truth for deployment-relevant timing, not
as a measurement of pure accelerator kernel latency. The 10 MS/s sample
rate sets a 100 ns floor on transition resolution, which is well below
the inference intervals of interest but matters for the marker-overlap
failure mode discussed in Section 3.2. The run 003 ambiguity discussed
in Section 4.4, Jetson-side vs.~host-side acquisition failure, is itself
a methodology limitation: a future iteration should instrument the
capture host independently to disambiguate.

\section{Conclusion}

We paired software-reported and externally observable timing for
MobileNetV2 inference on a Jetson Orin Nano across TensorRT and ONNX
Runtime, baseline and two interference conditions, with 3500 paired
samples across 35 capture runs and an additional 300 software-only
samples across 3 storage-stress runs where external pairing failed. The
two inference architectures differ in distributional structure as well
as mean latency, respond qualitatively differently to memory pressure
(tail inflation vs.~regime shift), and under storage writeback stress
diverge in three distinct ways that decouple inference execution
continuity from timing observability continuity. A single-timing-source
study would not have detected the decoupling. The general claim is
narrow: observability is a deployment-sensitive resource, and
characterizing edge inference under interference is more reliable when
the measurement infrastructure is itself independently observed.

\bibliography{references}

\end{document}